\documentclass[twocolumn,prb,superscriptaddress]{revtex4-2}
\usepackage{amsmath,amssymb,mathrsfs}
\usepackage{natbib}
\usepackage{subfigure}
\usepackage{tabularx}
\usepackage{epsfig}
\usepackage{longtable}
\usepackage{amsfonts}
\usepackage{rotating}
\usepackage{bbold}
\usepackage{hhline}
\usepackage{braket}
\usepackage{txfonts, comment}

\usepackage[colorlinks=true]{hyperref}

\hypersetup{linkcolor=magenta,urlcolor=blue,citecolor=blue,pdfstartview={FitH},hyperfootnotes=false,unicode=true}

\def\be{\begin{equation}}
\def\ee{\end{equation}}
\def\bea{\begin{eqnarray}}
\def\eea{\end{eqnarray}}
\def\nn{\nonumber}

\newcommand{\qdket}[2]{ {\left| \begin{array}{c} #1 \\ #2\end{array} \right\rangle}}  
\newcommand{\qdbra}[2]{ {\left\langle \begin{array}{c} #1 \\ #2\end{array} \right|}}

\begin{document}

\title{Peratic Phase Transition by Bulk-to-Surface Response}

\author{Xingze Qiu} 
\affiliation{State Key Laboratory of Surface Physics, Institute of Nanoelectronics and Quantum Computing, and Department of Physics, Fudan University, Shanghai 200438, China}
\affiliation{Shanghai Qi Zhi Institute, Shanghai 200030, China}
\author{Hai Wang} 
\affiliation{State Key Laboratory of Surface Physics, Institute of Nanoelectronics and Quantum Computing, and Department of Physics, Fudan University, Shanghai 200438, China}
\affiliation{Shanghai Qi Zhi Institute, Shanghai 200030, China}
\author{Wei Xia} 
\affiliation{State Key Laboratory of Surface Physics, Institute of Nanoelectronics and Quantum Computing, and Department of Physics, Fudan University, Shanghai 200438, China}
\author{Xiaopeng Li}  
\email{xiaopeng\_li@fudan.edu.cn}
\affiliation{State Key Laboratory of Surface Physics, Institute of Nanoelectronics and Quantum Computing, and Department of Physics, Fudan University, Shanghai 200438, China}
\affiliation{Shanghai Qi Zhi Institute, Shanghai 200030, China}
\affiliation{Shanghai Research Center for Quantum Sciences, Shanghai 201315, China}

\begin{abstract}

The study of dynamical phase transitions has been attracting considerable research efforts in the last decade. 
One theme of present interest is to search for exotic scenarios beyond the framework of equilibrium phase transitions. 
Here, we establish a duality between many-body dynamics and static Hamiltonian ground states for both classical and quantum systems. 
We construct frustration free Hamiltonians whose ground state phase transitions have rigorous duality to  chaotic transitions in dynamical systems. By this duality, we show the corresponding ground state phase transitions are characterized by bulk-to-surface response, which are then dubbed ``peratic" meaning defined by response to the boundary. For the classical system, we show how the time-like dimension emerges in the static ground states. For the quantum system, the ground state is a superposition of geometrical lines on a two dimensional array, which encode the dynamical Floquet evolution history of one dimensional disordered  spin chains. 
Our prediction of peratic phase transition  has direct consequences in quantum simulation platforms such as Rydberg atoms and superconducting qubits, as well as anisotropic spin glass materials. 
The discovery would shed light on the unification of dynamical phase transitions with equilibrium systems. 

\end{abstract}

\date{\today}

\maketitle

\section{Introduction}

Characterization of different phases of matter is fundamental to our understanding of nature.  With rapid developments in controlling many-body states away from equilibrium in condensed matter and quantum information experiments, the study of non-equilibrium phase transitions has attracted much present research interest. 
Non-equilibrium many-body physics previously unreachable can now be probed in experimental systems such as light-driven electronic matters~\cite{2020_Demsar_JLTP,2022_Rivera_NRP} and highly controllable quantum simulation~\cite{2021_Altman_PRXQ,2015_Bloch_Science,deng2016observation,smith2016many,2017_Lukin_Nature} or quantum computing platforms~\cite{2017_Monroe_TC,2017_Google_Science,2019_Pan_Science,2020_Wang_Science}. Quantum many-body localization (MBL) happens for a disorder system in its dynamical properties at infinite temperature~\cite{2015_Huse_Review,2019_Bloch_RMP}. Anomalous protected edge modes at zero Chern number which are unexpected for equilibrium systems, could appear in periodically driven Floquet quantum dynamics~\cite{2010_Demler_PRB,2011_Jiang_FloquetMajorana,2013_Levin_PRX}. Spontaneous translation symmetry breaking is generalized to the temporal domain, giving rise to crystallization in time ~\cite{2012_Wilczek_TC,2012_Wilczek_QTC,2016_Sondhi_PRL,2016_Nayak_PRL,2017_Monroe_TC,2017_Lukin_Nature}. There is mounting evidence  suggesting non-equilibrium phases may contain exotic scenarios  beyond equilibrium setups.

\begin{figure}[htp]
\centering
\includegraphics[width=0.48\textwidth]{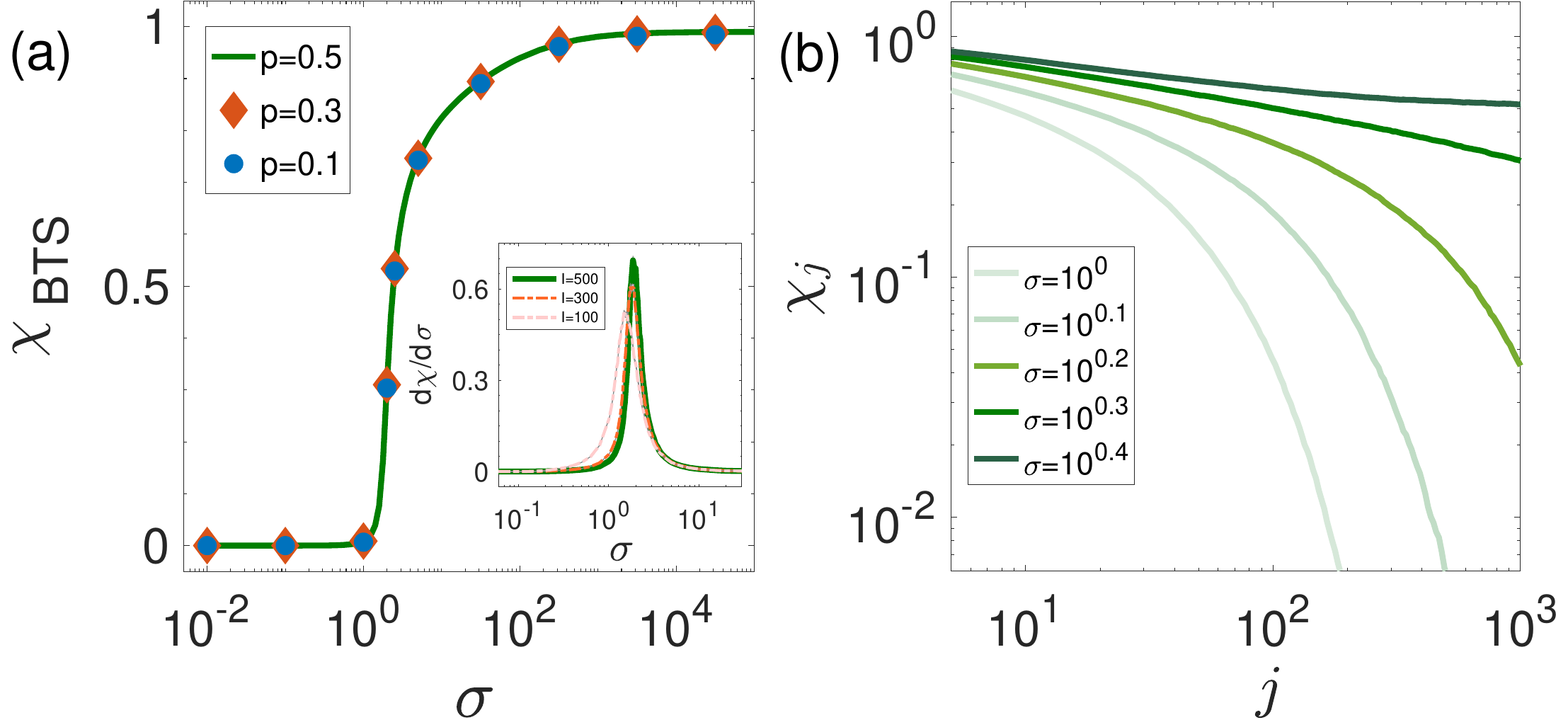}
\caption{Classical peratic phase transition. (a), the BTS response ($\chi_{\rm BTS}$) across the phase transition for {$I = 500$}. It vanishes at small coupling variance ($\sigma$) and becomes finite across the phase transition. The BTS response is averaged over random surface terms ($h_{\rm surf}$), with $p$ the probability of each surface term $h_{{\rm surf},i}$ taking a positive value. The inset of (a)  shows its first derivative for $I=100$, $300$, $500$ with $p=0.5$. This derivative develops a peak near the transition point, which sharpens up as we increase the system size. This implies a divergent second derivative in the thermodynamic limit. 
(b), the dependence of  local BTS response ($\chi_j$) on the $j$ index. This quantity has an exponential decay at small $\sigma$ when $\chi_{\rm BTS}$ vanishes. The decay becomes power law at the critical point ($\sigma_{\rm c}$). 
Here we choose $J=5I$ and $M = 2$.
The results are calculated by averaging over $10^4$ samples, and the statistical data error is smaller than the symbol size. 
}
\label{fig:classicalchi}
\end{figure}

Here, we construct frustration free Hamiltonians whose ground states have rigorous duality to dynamical systems. Through this theoretical construction, we find a novel phase transition mechanism---bulk-to-surface (BTS) response defines a peratic phase transition in Hamiltonian ground states. This mechanism is established by building exact duality  to order-to-chaos and MBL transitions for classical and quantum systems, respectively. 
The Hamiltonian  ground states have two distinctive phases as determined by whether the bulk is rigid against manipulations at the surface. 
The phase transition is characterized by a BTS response 
\be 
\chi_{\rm BTS} =  {\rm Var}  \left( \langle O_{\rm bulk} \rangle  \right)|_{\rm surface\, manipulation}     
\label{eq:chiBTS} 
\ee 
being zero or finite, 
with $\langle O_{\rm bulk}\rangle $ an observable in the bulk, ${\rm Var} (\langle O_{\rm bulk}\rangle ) $ the variance of the bulk observable as we manipulate the surface. The BTS response quantifies the stability of the bulk against surface manipulations, which vanishes if the bulk is stable and acquires a finite value otherwise. 
The phase transition in the classical ground state has an exact duality to the order-to-chaos transition in classical nonlinear dynamical systems~\cite{1988_Packard}. The quantum ground state phase transition constructed by forming superposition of fluctuating line configurations has a rigorous duality with the quantum ergodic~\cite{1991_Deutsch_PRA,1994_Srednicki_PRE,2008_Rigol_Nature} to MBL  transition~\cite{2015_Huse_Review,2019_Bloch_RMP} in quantum many-body dynamics. 
Unlike the standard phase transitions described either by spontaneous symmetry breaking~\cite{landau1937,Weinberg} or quantum state topology~\cite{2010_Kane_RMP,2011_Qi_RMP,2017_Wen_RMP}, there is no change in symmetry or topology across  the peratic phase transition. Our theory implies that the study of unconventional phases in non-equilibrium systems would rather inspire discovery of more exotic equilibrium phases and transitions than reaching completely beyond.

\section{  Emergent  chaotic dynamics in a Hamiltonian ground state } 

We first consider a discrete classical system which contains binary degrees of freedom $z_{ij} = \pm $ on a two dimensional (2d) grid, with the indices $i\in [0, I-1]$, and $j \in[ 0, J-1]$. Our theory starts from constructing a Hamiltonian ground state that supports the phase transition scenario described by the BTS response in Eq.~\eqref{eq:chiBTS}. The Hamiltonian contains a bulk and a surface term, $H_{\rm bulk}$, and $H_{\rm surf}$, 
\bea 
\label{eq:classicalHam}
H_{\rm bulk} &=& -\sum_{i} \sum_{j>0} { z_{ij} {\rm sgn} \left [ u_j+  \sum_{m=-M}^M w^{[ij]}_{\,m}  z_{i+m,j-1} \right]}\, , \nn  \\ 
H_{\rm surf} &=& -\sum_i h^{[i]}_{\rm surf} z_{i,0}\, ,  
\eea 
where $w^{[ij]}$ represents Ising couplings, $u_j$ is a local field in the bulk, and $h^{[i]} _{\rm surf}$ is a local field on the surface introduced to study the BTS response. 
The Ising couplings are randomly drawn from a Gaussian distribution with zero mean and variance $\sigma^2$. 
The bulk local field $u_j$ takes random binary values $\pm 1$ with equal probability. 
We adopt an open boundary condition along the $j$-axis and periodic boundary along the other axis (Fig.~\ref{fig:classicalchi}). 
The system has a layer structure along the $j$-axis---we have inter-layer couplings only.

Minimizing each term of the Hamiltonian in Eq.~\eqref{eq:classicalHam} leads to a set of equations, 
\be
\label{eq:classicalz}
z_{i0}^{\rm g}   = {\rm sgn} \left( h^{[i]} _{\rm surf} \right), \;
z_{i,j>0} ^{\rm g} = {\rm sgn} \left( u_j+ \sum_{m=-M}^M w^{[ij]}_{\, m}  z_{i+m,j-1} ^{\rm g}  \right)  , 
\ee
which can all be satisfied.  The Hamiltonian is thus frustration free. The binary degrees of freedom in the Hamiltonian ground state is given by $z_{ij}^{\rm g}$. 
We then observe that the $j$-dependence of the binary variables in the static ground state is time-like, for the $j$-th layer is completely determined by the $(j-1)$-th layer. A time-like dimension thus arises in the static ground state.  

By treating the $j$-axis as a time evolution direction, Eq.~\eqref{eq:classicalz} maps onto classical nonlinear dynamics having an order-to-chaos dynamical phase transition~\cite{1988_Packard}. 
In the ordered phase ($\sigma<\sigma_{\rm c}$), the dynamical state is pinned by the local field $u_j$---the difference between two initial states as measured by Hamming distance is quickly washed away in the dynamical evolution. 
In the chaotic phase ($\sigma>\sigma_{\rm c}$), the difference in the initial states  would either remain or gain amplification by the nonlinear dynamics.  Consequently, the ground state of the Hamiltonian in Eq.~\eqref{eq:classicalHam} also has two phases. For $\sigma<\sigma_{\rm c}$, the system is in a rigid phase where the bulk is stable and immune to surface manipulations with different $h_{\rm surf}$. For $\sigma>\sigma_{\rm c}$,  we have a volatile phase with the bulk sensitive to surface manipulations. Quantitatively, the BTS response for this specific system is defined as 
\bea 
\chi_{\rm BTS} = \lim_{h_{\rm surf}\to 0} \left[ \frac{1}{IJ} \sum_{ij} {\rm Var} \left( z_{ij}  \right)|_{h_{\rm surf}} \right], 
\eea 
with the variance ${\rm Var} \left( z_{ij}  \right)=1-\braket{z_{ij}}^2$, and the average $\braket{z_{ij}}$ obtained by randomly sampling $h_{\rm surf}$. 
This BTS response quantifies the degree of bulk fluctuations induced by perturbations at the boundary, and reflects the chaotic structures of the dual dynamics. 
We also introduce a space resolved local BTS response $\chi_j =I^{-1}  \sum_i {\rm Var}\left( z_{ij}  \right)|_{h_{\rm surf}}$ to diagnose the criticality.
Their behavior across the phase transition is shown in Fig.~\ref{fig:classicalchi}. The BTS response vanishes in the rigid phase, and becomes finite in the volatile phase, defining our peratic phase transition.   
The BTS response is consistent with the bulk entropy of the system (see Appendix \ref{sec:entropy}). 
By increasing the system size, we find the BTS response has a divergent second derivative at the critical point. 
The local BTS response shows an exponential decay in the rigid phase, with a decay length that tends to diverge approaching the critical point. At the critical point, the local BTS response exhibits a power law scaling, 
\be 
\textstyle \chi_j \propto j^{-\eta},
\ee   
a signature of nontrivial criticality for the peratic phase transition. 
We argue the critical behavior is described by a universal finite size scaling function 
\be 
\textstyle \chi_{\rm BTS} (t, L) = L^{-\eta} G(Lt^\nu)\, , 
\ee 
with $I, J \propto L$, $t= (\sigma-\sigma_{\rm c})/\sigma_{\rm c}$, and determine the anomalous dimension $\eta = 0.172(2)$, and $\nu$-exponent $\nu = 2.73(8)$ (see Appendix \ref{sec:FSS}). 
This phase transition scenario does not rely on the symmetry or the topology of the system, in sharp contrast to the standard phase transitions. 
Whether it can be described by the replica type of spontaneous symmetry breaking~\cite{SimonsBook,Edwards_1975,1979_Parisi_PRL} as established in disorder spin systems~\cite{1986_Binder_RMP,2017_Pierangeli_NC,2018_Dwave_Science} is worth future investigation. We emphasize that the chaotic structures or the BTS response reported here for the  highly anisotropic disorder spin system may shed light on understanding anisotropic  spin glasses~\cite{1986_Binder_RMP,1979_Wanklyn_PRL,2021_Greedan_PRR},  for which  the exact models are lacking to our knowledge.

\begin{figure}[htp]
\centering
\includegraphics[width=.48\textwidth]{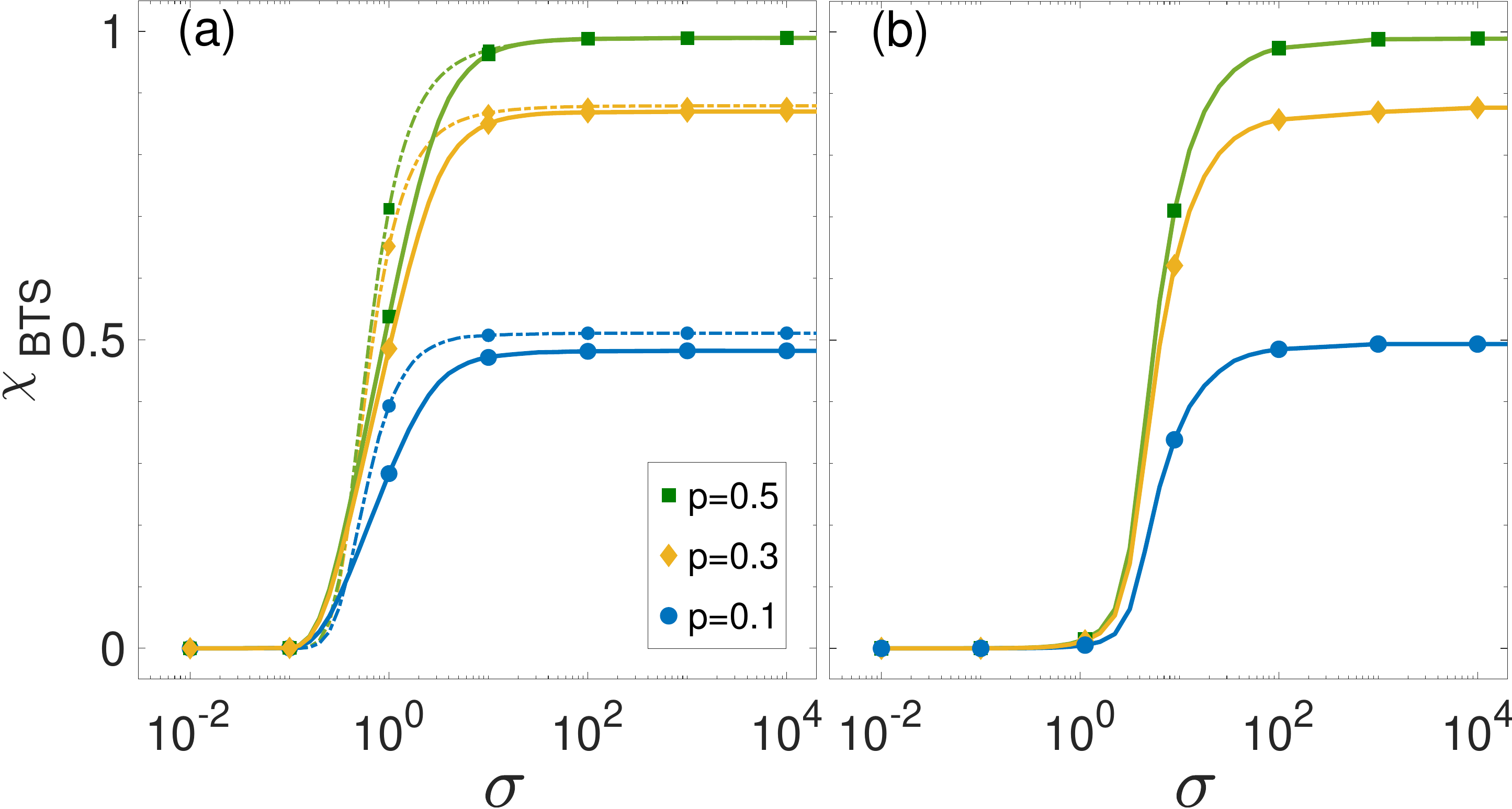}
\caption{
The peratic phase transition in the frustrated Ising model (Eq.~\eqref{eq:frustratedmodel}). 
(a), the classical peratic phase transition. The dashed lines show the results of the frustration free model for comparison.
(b), the quantum peratic phase transition of the frustrated Ising model adding the transverse field ($h_{\rm T}$). 
We choose a field strength $h_{\rm T} = 10$. 
The BTS response is averaged over different surface terms with $p=0.1, 0.3, 0.5$. In this plot, we choose $I=5$, $J=5$, and $M = 2$. 
The results are calculated by averaging over $10^4$ samples, and the statistical data error is smaller than the symbol size. 
}
\label{fig:IsingZX}
\end{figure}

One fascinating property of the volatile phase is its robustness---the BTS response is stable irrespective of different choices of surface manipulations. In our numerical calculation, each surface term $h_{ {\rm surf}, i}$ takes a positive value with a probability $p$, and a negative value with  $1-p$. The BTS response in the volatile phase as constructed above does not vary with the $p$-value (Fig.~\ref{fig:classicalchi}(b)), i.e., having robustness against different surface manipulations. This  nontrivial property can be attributed to the presence of a dynamical fixed point of the Hamming distance evolution in the dual chaotic dynamics~\cite{2004_Bertschinger_NC}. This  makes the volatile phase sharply distinctive from a trivial case with the bulk trivially determined by the surface, for example with $z_{ij} = z_{i0}$, where the BTS response would strongly depend on $p$.

\section{  Experimental candidates }

Although the frustration free Hamiltonian in Eq.~\eqref{eq:classicalHam} has a nice property of being polynomially solvable, its experimental realization is challenging. For experimental realization, we further consider a frustrated Hamiltonian with two-body Ising couplings only, 
\be 
H_{\rm bulk} = -\sum_{i} \sum_{j>0} { z_{ij} \left [ u_j+  \sum_{m=-M}^M w^{[ij]}_{m}  z_{i+m,j-1} \right]}, 
\label{eq:frustratedmodel}
\ee 
which could describe a broad range of spin systems from Rydberg atomic systems~\cite{2010_Saffman_RMP} and superconducting qubits~\cite{2020_Oliver_ARCMP} to anisotropic spin glasses~\cite{1986_Binder_RMP}. 
We study its ground state phase transition by numerically minimizing the energy. The results are shown in Fig.~\ref{fig:IsingZX} (a). We observe that the BTS responses for the frustrated and frustration-free models are quite similar to each other with a tiny difference barely noticeable. The peratic phase transition still preserves in the frustrated model.

\begin{figure}[htp]
\centering
\includegraphics[width=0.48\textwidth]{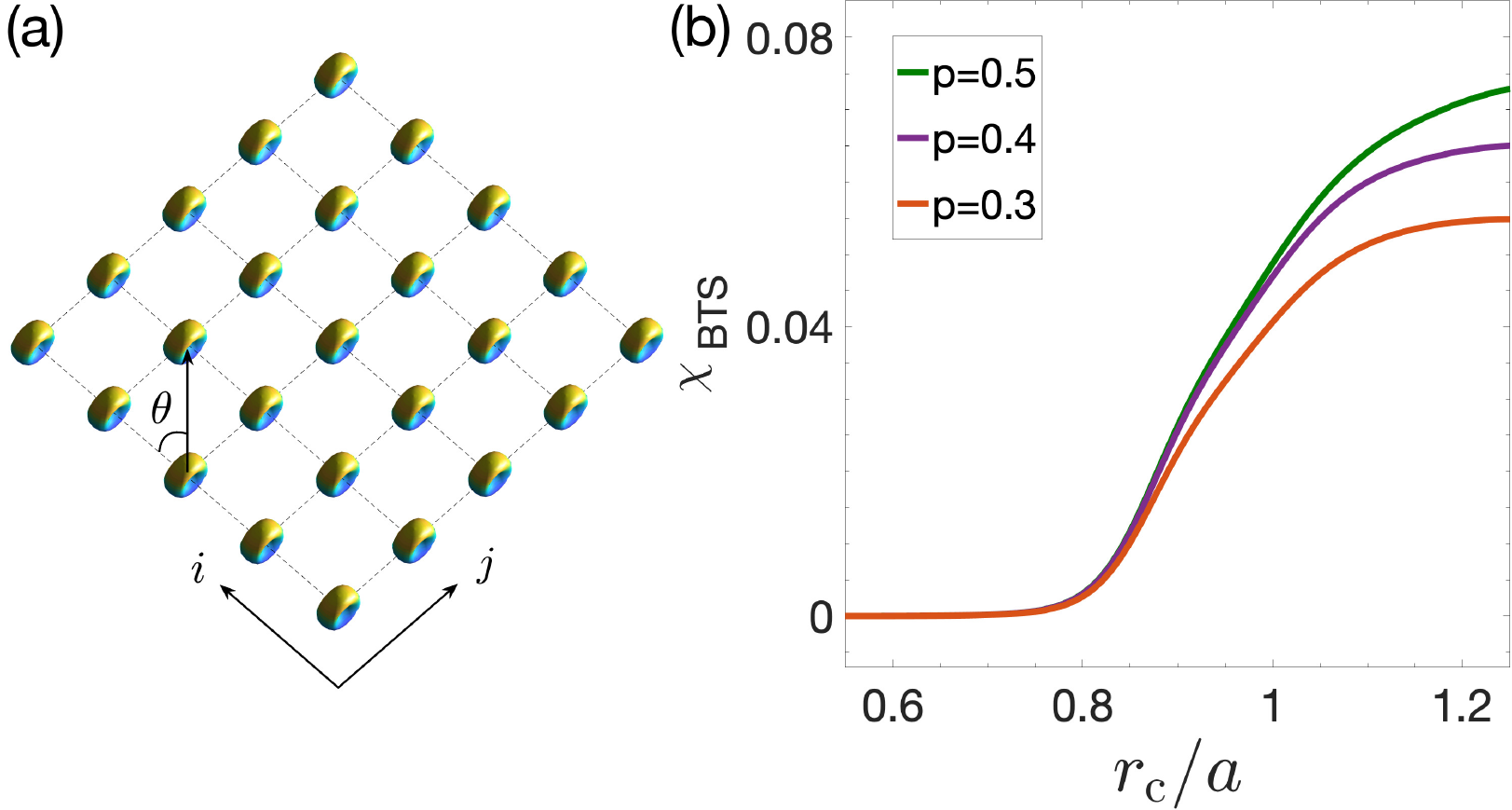}
\caption{
Peratic phase transition with Rydberg atom arrays. 
(a), schematic illustration of the Rydberg system. The atoms are dressed to a Rydberg $p$-wave state and located on a 2D square lattice, whose lattice constant is $a$. 
(b), the BTS response across the peratic phase transition. 
The strength of longitudinal fields are fixed along the $i$-axis, and random drawn from the binary values $\pm V$ with equal probability along the $j$-axis. 
The BTS response is averaged over all configurations of the different longitudinal fields and different surface terms, with $p=0.3, 0.4, 0.5$. 
Here, we set the transverse field strength $h_{\rm T} = 0.5V$, and the system size as $5\times5$. 
The results are calculated by averaging over all possible surface spin polarizations. 
}
\label{fig:Rydberg} 
\end{figure}

The two-body Hamiltonian also allows a natural way to generalize the phase transition to quantum systems, by simply promoting the Ising variables $z_{ij}$ to Pauli operators $\hat{z}_{ij}$, and adding a transverse field coupling $\Delta H  = \sum_{ij} h_{\rm T} \hat{x}_{ij} $ ($\hat{x}$ the Pauli-x operator) to generate quantum fluctuations. As shown in Fig.~\ref{fig:IsingZX} (b), the peratic phase transition still persists in presence of quantum fluctuations. We find quantum effects further stabilize the rigid phase and shift the transition point towards the volatile phase, which is somewhat counterintuitive. This can be attributed to that the transverse field couples the degenerate ground states and develops a tendency for gap opening, rendering the bulk more rigid.

\begin{figure*}[htp]
\centering
\includegraphics[width=0.99\textwidth]{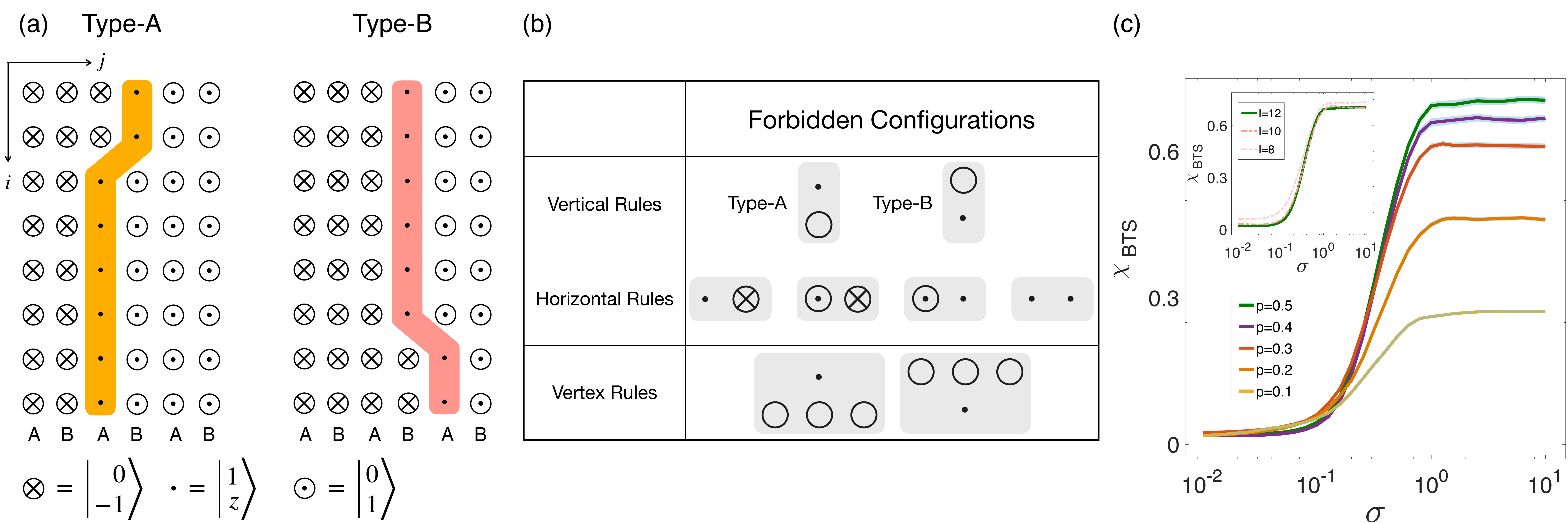}
\caption{
Quantum peratic phase transition. 
(a), 
schematic illustration of the frustration free qudit model. The model contains a 2d array of qudits, with its four states marked by `$\cdot$', and `$\bigcirc$'. The two `$\bigcirc$' states are further specified as  `$\odot$', and `$\otimes$'. (b), the forbidden configurations. The corresponding projector Hamiltonian is provided in Appendix~\ref{sec:QHam} . The consequent  low-energy subspace corresponds to the two types of line configurations as illustrated in (a) ({\it see main text}). 
(c), the BTS response across the quantum peratic phase transition. Here we sample the surface terms $h_{\rm surf}$ with different $p$ values from $0.1$ to $0.5$. 
The quantum ground state phase transition as constructed is dual to the quantum ergodic to many-body-localized dynamical phase transition.  At small $\sigma$, the bulk is robust against surface manipulations giving a vanishing BTS, which is dual to the dynamical quantum ergodic phase. Above a certain threshold at $\sigma>\sigma_{\rm c} $, the bulk is stringently tied with surface, corresponding to the dynamical MBL phase. 
The results are calculated by averaging over 100 samples, with the standard deviations illustrated by the shaded error bands.
The inset of (c) shows the system size dependence of BTS response (with $p$ fixed at $0.5$), which indicates the phase transition becomes sharper at larger system size. 
}
\label{fig:MBLErgodic}
\end{figure*}

As a concrete experimental candidate, we consider a system of Rydberg $p$-wave dressed atoms, which has been used to construct quantum spin ice Hamiltonians~\cite{2014_Glaetzle_PRX} and programmable quantum annealing~\cite{2020_Qiu_PRXQ}.
Two atomic hyperfine states 
$\ket{\uparrow} = \ket{5^2 S_{1/2}, F = 2, m_F = 0}$ 
and 
$\ket{\downarrow} = \ket{5^2 S_{1/2}, F= 1, m_F = 0}$ 
of $^{87}$Rb atoms are selected to form a spin-$1/2$ lattice with $I$ rows and $J$ columns. 
Their couplings are controllable by performing a microwave induced transition, which are described by a transverse field 
$ 
 H_{\rm T} = \sum_{{\bf r}} h_{\rm T} \hat{x} _{{\bf r}}, 
$  
with the field strength $h_{\rm T}$ determined by the Rabbi frequency of the microwave. 
Taking a Rydberg $p$-wave dressing scheme where the $\ket{\uparrow}$ state is selectively dressed with a Rydberg $p$-wave state $|n^2 P_{3/2}, m= 3/2\rangle$ (the quantization axis along the $i$-direction in Fig.~\ref{fig:Rydberg} (a)), we introduce interactions between neighboring layers as labeled by $j$~\cite{2014_Glaetzle_PRX}. 
One key feature of this system is it has inter-layer interactions but no intra-layer interactions, and thus the description of this system closely resembles Eq.~\eqref{eq:frustratedmodel}.
Arranging the atoms periodically in a 2D array as shown in Fig.~\ref{fig:Rydberg}, 
the resultant interaction between the two qubits at ${\bf r}$ and ${\bf r}'$ is given by 
\be 
\label{eq:Rydberg} 
H_{\rm R}  = \sum_{{\bf r}, {\bf r}'} 
\frac{V\sin^4 \theta_{{\bf r}{\bf r}'} }{1+ \left ({ |{\bf r}-{\bf r}' |}/{r_{\rm c}} \right )^6 } (\hat{z}_{\bf r} +1) (\hat{z}_{{\bf r}'}+1) /2 , 
\ee 
with the coupling strength $V$ and the Rydberg interaction range $r_{\rm c}$ determined by the Rabbi frequency and the detuning of the one-photon transition of the Rydberg dressing scheme~\cite{2014_Glaetzle_PRX}.  
The on-site  longitudinal  field is tunable by  manipulating the detuning of the microwave with respect to the hyperfine splitting of $6.8$ GHz. The longitudinal fields are described by 
$
H_{\rm L} = \sum_{{\bf r}} h_{{\bf r}}  \hat{z}_{{\bf r}}. 
$

As shown in Fig.~\ref{fig:Rydberg} (b), the BTS response shows distinctive behaviors at small and large Rydberg interaction range $r_{\rm c}$. 
At small $r_{\rm c}$, the system is in a  rigid phase having a vanishing $\chi_{\rm BTS}$ with bulk spin polarization robust against 
different choices of surface polarizations. When $r_{\rm c}$ is larger than a certain threshold (roughly $0.8$ in units of lattice constant), the bulk becomes volatile, fluctuating with different  surface polarizations, as characterized by a finite $\chi_{\rm BTS}$. The Rydberg system thus supports a peratic phase transition characterized by the BTS response. 
Since two spin states in our proposed setup has a hyperfine splitting at the order of GHz, the peratic phase transition can be detected by microwave spectroscopy, which is a standard technique in cold atom experiments.  The local addressability can be reached by creating spatial-resolved Stark shifts using focused laser fields~\cite{Weiss2016Science}.

\section{ Quantum peratic phase transition }  
We now provide a rigorous quantum  model supporting the quantum peratic phase transition. 
This is achieved by constructing a quantum frustration free Hamiltonian, whose ground state phase transition has an exact duality with the dynamical MBL transition of the Floquet quantum dynamics of a one-dimensional disordered spin chain~\cite{2015_Huse_Review,2019_Bloch_RMP}.

We consider a 2d qudit lattice model with a four dimensional local Hilbert space. The four levels are labeled as 
$\left| 
\begin{array}{c} 
\nu \\ 
z 
\end{array} 
\right \rangle$ 
with $\nu$ ($= 0, 1$) and $z$ ($=\pm 1$) (Fig.~\ref{fig:MBLErgodic} (a)). 
The lattice contains $I$ rows and $J$ rungs, with two types of rungs labeled by $A$ and $B$. 
Different qudits are labeled according to their position on the lattice by $(i, j)$, with $i$ ($j$) the row (rung) index. 
The large $4^{IJ}$-dimensional Hilbert space of the qudit system is constrained to a low-energy subspace by introducing local projectors. The forbidden configurations are illustrated in Fig.~\ref{fig:MBLErgodic} (b), with the corresponding Hamiltonian realization given in Appendix~\ref{sec:QHam}. With the forbidden configurations by the horizontal rules, there is at most only one `$\cdot$' in one row. All sites on the left (right) of the `$\cdot$' have to be `$\otimes$' (`$\odot$'). 
%All sites on its left have to be . 
By the vertex rules, the `$\cdot$' sites on each rung have to form a continuous line, which can either go straight on the same rung or bend over to its nearby rungs.  By the vertical rules on the A (B) rungs, the continuous line has to reach to the bottom (top)  on the A (B) rungs. With all these constraints, the allowed states in the low-energy subspace correspond to the two types of continuous lines---Type $A$ and Type $B$, as illustrated in Fig.~\ref{fig:MBLErgodic} (a). 
Type $A$ ($B$) line starts from the bottom (top) at $A$ ($B$) sites, goes straight upward (downward), bends at most once to the right, and then continues upward (downward) following the $B$ ($A$) rungs to the top (bottom). 
There are a total number of $N = I(J-1)+1$ such lines, with each line uniquely labeled by $(i,j)$, according to the site where the line bends over, or the last site if the line does not bend. 
The quantum state on the line is 
$\left| 
\begin{array}{cccc} 
1   &1  &\ldots     &1 \\ 
z_0 &z_1 &\ldots    &z_{I-1}
\end{array} 
\right\rangle $, with the subscripts labeled by the order from top to bottom on the lattice. 
The qudits to its left (right) all reside on the state $\qdket{0}{-1}$ $\left(\qdket{0}{1}\right)$. 
All the low-energy states can then be labeled as $|i,j; \vec{z} \rangle$ [$\vec{z} \equiv (z_0, z_1 \ldots z_{I-1} )$].

We construct a 2d local qudit Hamiltonian (see Appendix \ref{sec:QHam}), whose projection on the low-energy subspace takes a form of Feynman-Kitaev clock Hamiltonian~\cite{2002_Kitaev_Book,2007_Lloyd_SIAM}, 
\bea 
\label{eq:HamEff} 
&& H_{\rm eff} = \sum_{\vec{z}_\star}  \left\{ 
-\frac{1}{2} \left( |\gamma_0 (\vec{z}_\star) \rangle \langle \gamma_0 (\vec{z}_\star) | 
    -  |\gamma_{N-1} (\vec{z}_\star ) \rangle \langle \gamma_{N-1} (\vec{z}_\star) |  \right)  \right.   \nn \\
   && + \sum_{l=0}^{N-2} \left.  \frac{1}{2}  \left(  |\gamma_l (\vec{z}_\star) \rangle \langle \gamma_l (\vec{z}_\star) | 
                        - |\gamma_l (\vec{z}_\star)\rangle \langle \gamma_{l+1} (\vec{z}_\star) | +H.c. \right) \right\}.     
\eea 
Here, the $\gamma$-states are 
$
|\gamma_l (\vec{z}_\star) \rangle = \sum_{\vec{z}} \psi_l( {\vec{z}} )  |l; \vec{z} \rangle, 
$
and the sequential index, $l$, is introduced for a compact representation of  $(i,j)$---$l_{ij} = jI+i$ for $j\in A$, and $l_{ij} = (j+1)I-i-1$ for $j\in B$ 
(see one explicit example in Appendix~\ref{sec:QHam}, Fig.~\ref{fig:suppfigHistoryStates}).

The wave functions $\psi_l(\vec{z})$ are defined through a sequential unitary transformation starting from 
$\psi_0 (\vec{z}) = \delta_{\vec{z} \vec{z}_\star }$.  
The update of $\psi_l$ is designed to follow the Floquet quantum dynamics of a one-dimensional spin-$1/2$ system~\cite{2015_Abanin_PRL}. 
From $l=0$ to $l=I-1$, the update of $\psi_l$ corresponds to a unitary gate ${\rm e}^{-{\rm i}\hat{x}_i \hat{x}_{i+1}/10}$ ($i$ from $0$ to $I-1$), and these unitary gates are then applied in a backward order from $l=I$ to $2I-1$. Then in the same order, we apply the unitary gates  ${\rm e}^{-{\rm i}\hat{y}_i \hat{y}_{i+1}/10 }$ from $l=2I$ to $l=4I-1$, and ${\rm e}^{-{\rm i} \left(\hat{z}_i \hat{z}_{i+1}/10+\delta_i\hat{z}_i\right)}$ from $l=4I$ to $6I-1$. 
This unitary update process is then repeated forward for $J/6$ periods. 
The amplitude $\delta_i$ is a random number drawn from a Gaussian distribution with zero mean and variance $\sigma^2$. 

The constructed 2d local qudit Hamiltonian is semi-positive definite and frustration free~\cite{2002_Kitaev_Book}, whose ground state is 
$
|G (\vec{z}_\star)  \rangle = \frac{1}{\sqrt{N}}  \sum_l | \gamma_l (\vec{z}_\star) \rangle, 
$
having a $2^I$-fold degeneracy as labeled by $\vec{z}_\star$. The ground state is an equal-amplitude quantum superposition of those geometrical line configurations in Fig.~\ref{fig:MBLErgodic} (a).

We introduce a bulk observable to diagnose the peratic phase transition, 
\be 
\textstyle {O}_{ij}  = \textstyle \qdket{1}{1} \qdbra{1}{1} - \qdket{1}{-1} \qdbra{1}{-1}\, , 
\ee 
that acts on the site $(i,j)$. 
The ground state degeneracy would be lifted up by adding a perturbation on the edge 
$ \Delta H = \sum_i h^{[i]}_{\rm surf} [{1} - O_{i,0}]$---different choices of $h_{\rm surf}$ select ground states  $|G(\vec{z}_\star)\rangle$ with different  $\vec{z}_\star$. 
The corresponding BTS response is  
\[
\textstyle \chi_{\rm BTS} = \lim_{h_{\rm surf}\to 0} 
\left[ \frac{1}{I} \sum_{ij} \left ( \overline{ \langle O_{ ij}   \rangle^2}  - \overline{ \langle O_{ij} \rangle}^2  \right)|_{h_{\rm surf}} \right],
\] 
given by 
$
\chi _{\rm BTS} = \frac{1}{JI^2} \sum_{i,l} \left\{  \overline{\left(\sum_{\vec{z}} {z_i |\psi_l(\vec{z})|^2}\right)^2} - \overline{\left(\sum_{\vec{z}} {z_i |\psi_l(\vec{z})|^2}\right)}^2  \right\} , 
$
with $\overline{\ldots}$ averaging over different $\vec{z}_\star$. 
Treating unitary update from $l$ to $l+1$ as  quantum time evolution, the dynamics of $\psi_l(\vec{z})$ 
 has two distinctive phases---quantum ergodic and MBL. 
For the MBL dynamics, the wave function holds the memory of the initial configuration of  $\vec{z}_\star$, and the resultant BTS response $\chi_{\rm BTS}$ of the dual 2d quantum ground state is finite, 
{and strongly depends on $\vec{z}_\star$, namely the $p$-value in sampling $h^{[i] }_{\rm surf}$.} On the contrary for the quantum ergodic dynamics, the local observables would thermalize as $l$ proceeds and are then independent of $\vec{z}_\star$, rendering a vanishing BTS response for the 2d ground state.  The numerical results are provided in Fig.~\ref{fig:MBLErgodic} (c), which agree well with the theoretical analysis. The BTS response thus defines a peratic quantum phase transition in the frustration free quantum ground states.
Since the existence of the MBL phase has been proven for one-dimensional systems by Imbrie~\cite{2016_Imbrie_MBL}, 
our constructed exact duality establishes a rigorous scenario for the quantum peratic phase transition.

With the exact construction presented above and the numerical results for the transverse field Ising model, we expect the quantum peratic phase transition to be generic, not relying on the duality with the MBL to ergodic phase transition. The unconventional  phase transition defined  by bulk-to-surface response could arise in a broad range of quantum simulation platforms as well as anisotropic spin glass materials.

\section{ Conclusion and Outlook }
We propose  a peratic phase transition defined by a bulk-to-surface response in both classical and quantum ground states. By constructing frustration free models, we establish rigorous duality from the peratic phase transition to order-to-chaos transition in the classical system, and to MBL-to-ergodic transition in the quantum setting. 
With numerical results, we show the peratic phase transition also preserves in anisotropic spin glass models with two-body Ising couplings only. We predict the system of Rydberg p-wave dressed atoms supports the peratic phase transition. 
Our theory implies dynamical phase transitions would inspire exotic scenarios in equilibrium systems rather than reaching beyond. 
Our approach also provides an alternative way for characterizing dynamical phase transitions from the perspective of equilibrium phase transitions, which could unify the description of non-equilibrium phases within the equilibrium framework, for the constructed duality from dynamical phases to static Hamiltonian ground states is quite generic (see Appendix \ref{sec:DynamicMapping}).

\begin{acknowledgments}

We would like to thank Wei Wang for suggesting the ancient Greek ``p\'eras" for naming the phase transition, and thank Dong-Ling Deng and Meng Cheng for helpful discussion. 
This work is supported by National Program on Key Basic Research Project of China (Grant No. 2021YFA1400900, 2017YFA0304204),  National Natural Science Foundation of China (Grants No. 11774067, and 11934002), Shanghai Municipal Science and Technology Major Project (Grant No. 2019SHZDZX01), Shanghai Science Foundation (Grant No. 21QA1400500, 19ZR1471500). 
Xingze Qiu acknowledges support from National Natural Science Foundation of China (Grants No. 12104098).
\end{acknowledgments}

\appendix

\begin{figure}[htp]
\centering
\includegraphics[width=0.35\textwidth]{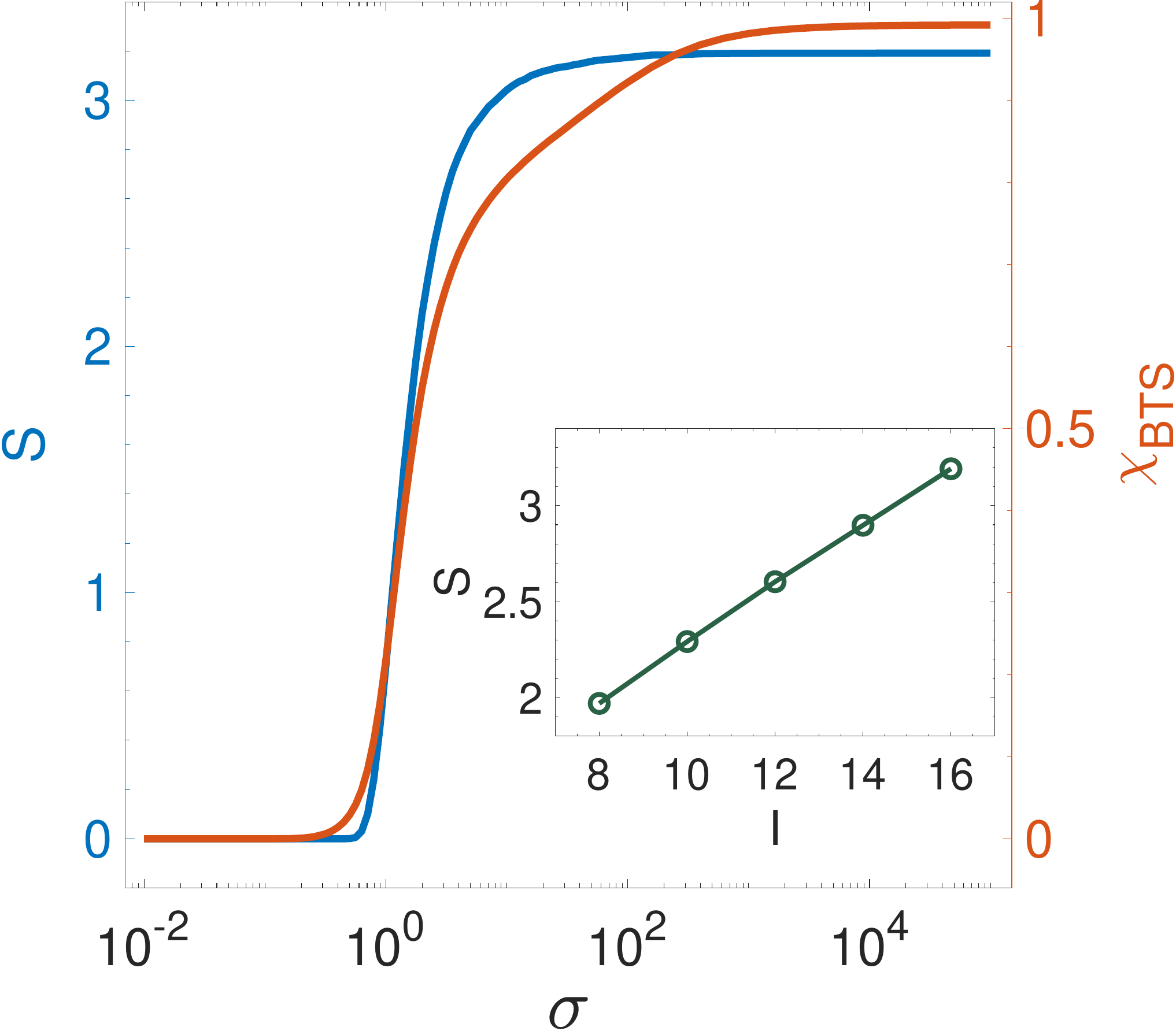}
\caption{
Phase transition in the frustration free Ising model (see Eq.~\eqref{eq:classicalHam} in the main text). Both the half-system entropy and the BTS response can determine two distinctive phases. The inset shows the linear scaling of the  entropy with the system size $I$, where we set $\sigma=10^4$. The results are averaged over $5\times{10}^4$ random samples. Here, we choose $I=16$, $J=5I$, $p=0.5$, and $M=2$ in this plot. 
}
\label{fig:entropy}
\end{figure}

\section{Relation between bulk-to-surface response and bulk fluctuations}
\label{sec:entropy}

In this section, we show that the bulk-to-surface (BTS) response reflects the entropy of bulk fluctuations in the system. 
To probe the bulk phase transition directly in our model, we define a half-system entropy $S=-\sum_{\boldsymbol \sigma}{P({\boldsymbol \sigma})}\log_2{P({\boldsymbol \sigma})}$, with ${\boldsymbol \sigma}$ indexing the configuration of one half of the system and $P({\boldsymbol \sigma})$ the corresponding probability. Its behavior is shown in Fig. \ref{fig:entropy}. We find that the half-system entropy is vanishing and sub-extensive on the two sides of the peratic  phase transition. 
Here, sub-extensive means that the entropy increases linearly with the system size $I$ instead of  $I\times J$. 
Both of the BTS response and the half-system entropy characterize the fluctuations of the bulk system, and serve properly as an order parameter for the peratic phase transition. But for our frustration-free model, it is so much more convenient to compute the BTS response than the entropy. Computing the BTS response takes polynomial time, whereas computing the entropy scales exponentially. The BTS response is thus more convenient to diagnose peratic phase transition.

\section{Phenomenological theory and scaling analysis} 
\label{sec:FSS}

To analyze the peratic phase transition, we propose a phenomenological theory and perform scaling analysis. Since the BTS response that characterizes the phase transition also probes the degree of fluctuations in the bulk system, resembling the entropy, the phenomenological  free energy near the phase transition takes a form of 
\be 
f (t, \chi_{\rm BTS} ) = -r(t) \chi_{\rm BTS} + \kappa (t) \chi_{\rm BTS}^2 + {\cal O}(\chi_{\rm BTS} ^3 ) ,  
\ee 
where we have $t = (\sigma-\sigma_c)/\sigma_c$, $r>0$ ($r<0$) for $t>0$ ($t<0$), and $\kappa>0$. This free energy does not have Ising symmetry. 
Nonetheless, minimizing the free energy under the physical constraint $\chi_{\rm BTS} \geqslant 0$ produces a phase transition, i.e., the peratic phase transition established in the main text.

\begin{figure}[htp]
\centering
\includegraphics[width=0.48\textwidth]{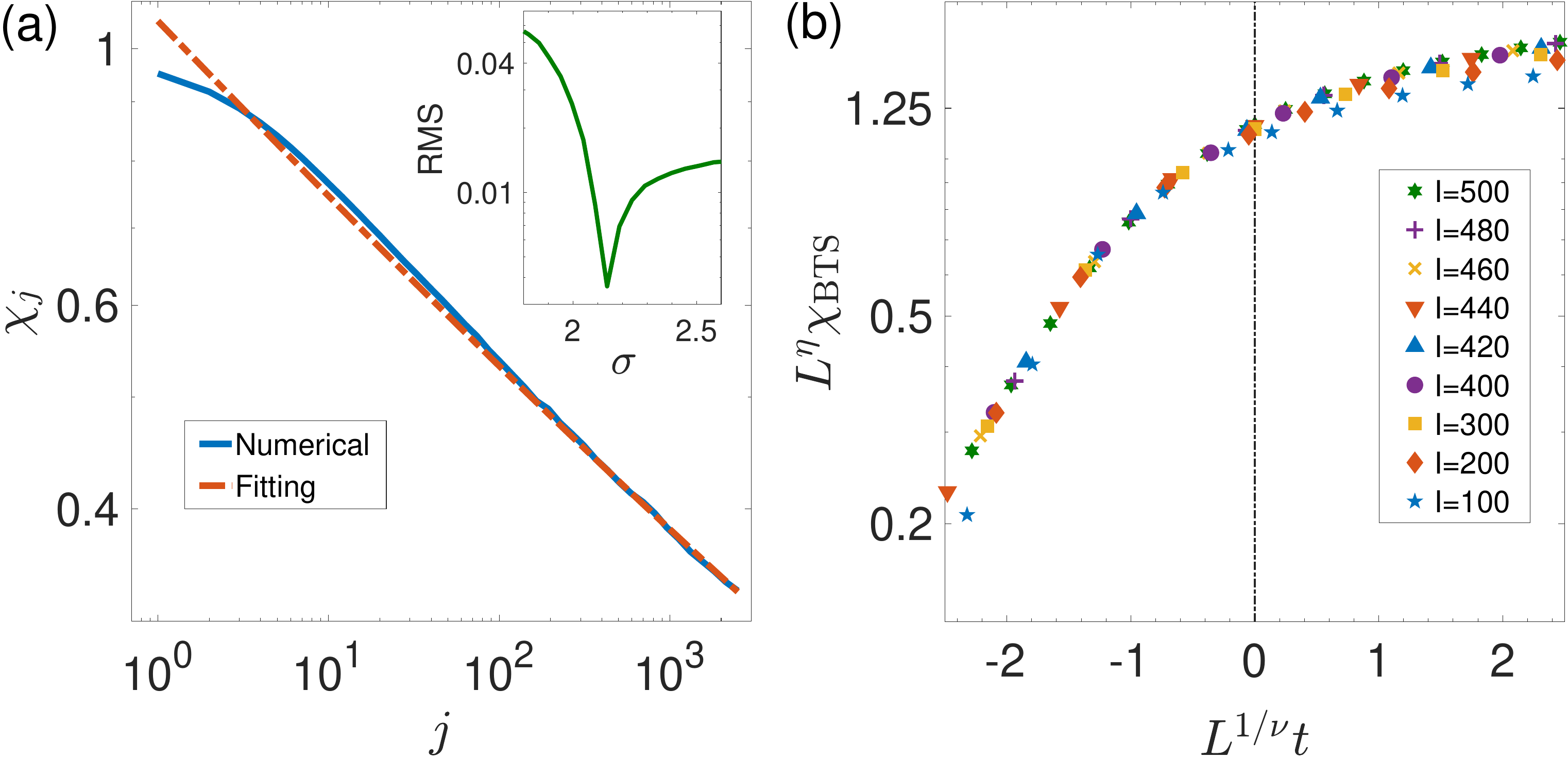}
\caption{
Finite size scaling analysis for the classical peratic phase transition. 
(a), power law decay of the local bulk-to-surface (BTS) response in $\chi_j$ and the $\eta$-exponent.  This exponent is obtained by fitting our numerical results for $\chi_j$ to a power law at the critical point. The inset shows the RMS error for fitting $\chi_j$ near the critical point to a power law function, and the location of the minimum indicates the critical point. In this plot, we choose $I=500$, $J=5I$, $p=0.5$, and $M=2$.  The $\eta$ exponent is obtained to be $0.172(2)$. 
(b), data collapse in the scaling analysis for the BTS response. Here, we choose $M=2$, $p=0.5$, $L=I$, $J=5I$. In (b), we take $\nu = 2.732$, which gives the best-quality data collapse. 
}
\label{fig:Criticality}
\end{figure}

The scaling analysis is performed by postulating the phase transition is described by a certain renormalization group fixed point~\cite{SimonsBook}. Taking a renormalization group transformation, $t\to t\zeta ^{1/\nu}$, $L\to L/\zeta $ (the system size $I,J\propto L$), with $\zeta$ a scaling factor, and $\nu$ a critical exponent, we have 
\be 
\chi_{j/\zeta} (t\zeta^{1/\nu} , L/\zeta) = \zeta^{\eta} \chi_{j} (t, L) , 
\ee 
at the neighborhood of the fixed point. Here, $\eta$ is the anomalous dimension.  The $\nu$ and $\eta$ exponents are to be fixed with our numerical results. 
The scaling form implies that 
\be 
\chi_j(t,L) = j^{-\eta} A(jt^\nu, Lt^\nu), 
\ee 
with $A(\ldots)$ a universal function. 
It then follows directly that a thermodynamic limit system right at the peratic phase transition point has a power law local BTS response, 
\be 
\chi_j \propto j^{-\eta}. 
\ee 
The scaling of the total BTS response obeys  
\be 
\chi_{\rm BTS} (t\zeta^{1/\nu}, L/\zeta) = \zeta^{\eta} \chi_{\rm BTS} (t, L), 
\ee 
which implies 
\be
\chi_{\rm BTS} (t, L) = L^{-\eta} G(L t^\nu ), 
\label{eq:BTSscaling}
\ee 
with $G(\ldots)$ a universal function.

The $\eta$-exponent is determined by fitting our numerical results to $\chi_j \propto j^{-\eta} $ at the critical point (Fig.~\ref{fig:Criticality} (a)), from which we get $\eta = 0.172(2)$. 
We first fit all numerical results near the critical point for $\chi_j$ to a power law, and locate the critical point by minimizing the root mean square (RMS) error of the fitting. 
The value of the $\eta$-exponent and its error are obtained by fitting the data right at the critical point.

The $\nu$-exponent is extracted by performing a data collapse taking the scaling form in Eq.~\eqref{eq:BTSscaling} (Fig.~\ref{fig:Criticality} (b)), and we get 
{
$\nu = 2.73(8)$. The $\nu$ exponent is calculated using the analysis in Ref.~\cite{2001_Bhattacharjee}, with the error estimated by bootstrap.}

\begin{figure*}[htp]
\centering
\vspace{1cm} 
\includegraphics[width=0.99\textwidth]{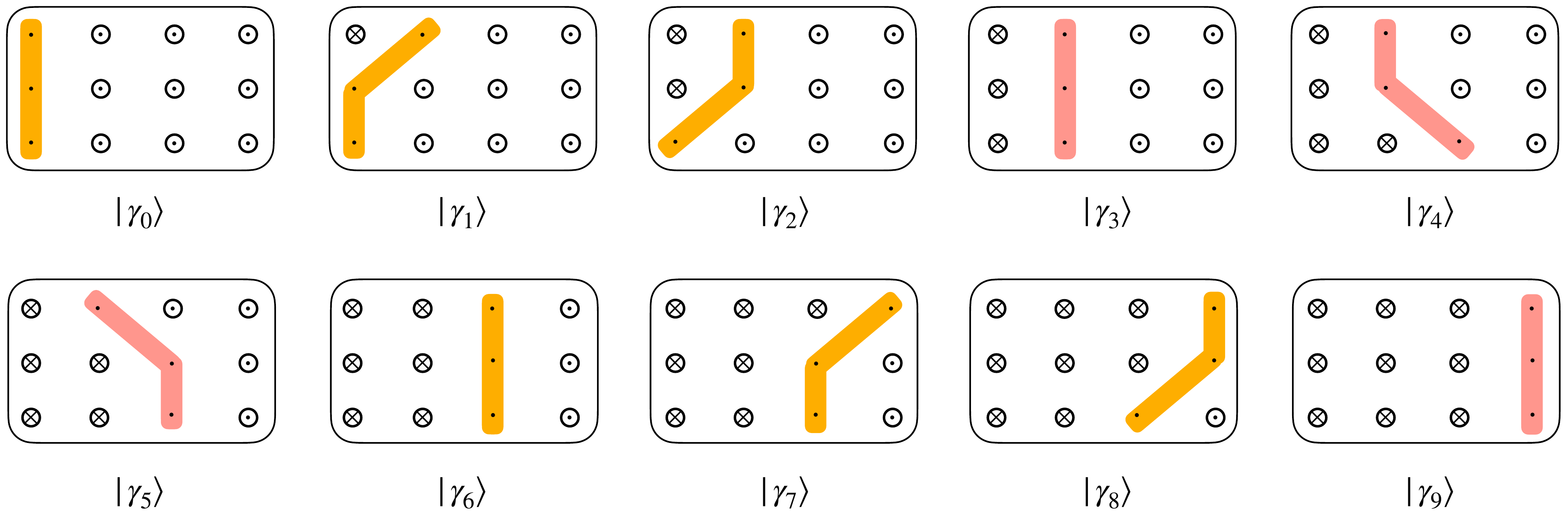}
\caption{ 
Line configurations for the quantum frustration free qudit model.  These line configurations represent the allowed states in the low energy subspace after enforcing the local projectors. Here we choose $I = 3$, and $J =4$ for illustration. 
}
\label{fig:suppfigHistoryStates}
\end{figure*}

\section{Construction of a frustration free quantum model} 
\label{sec:QHam}

We consider a 2d array of qudits. Its Hilbert space is projected to a low-energy subspace by introducing an energy penalty 
\[ 
H_P = \lambda \left( P_1 + P_2 + P_3+P_4 \right),
\] 
with $P_{1,2,3,4}$ four local projectors as described below. The projector $P_1$ acts on columnwise nearby sites $(i,j)$, and $(i+1,j)$ ({\it see Fig.~\ref{fig:MBLErgodic} in the main text for illustration}), 
\[
P_1 = \sum_{zz'} \left[ \sum_{i,j\in A} 
{\textstyle 
\left| 
	\begin{array} {cc} 
	1 	&0 	\\
	z 	&z'	
	\end{array} 
\right\rangle 
\left \langle
	\begin{array} {cc} 
	1 	&0 	\\
	z 	&z'	
	\end{array} 
\right| 
} 
+ \sum_{i,j\in B} 
{\textstyle \left| 
	\begin{array} {cc} 
	0 	&1 	\\
	z 	&z'	
	\end{array} 
\right\rangle 
\left \langle
	\begin{array} {cc} 
	0 	&1 	\\
	z 	&z'	
	\end{array} 
\right| 
}
\right]. 
\]
The projectors $P_2$ an $P_3$ both act on rowwise nearby sites $(i,j)$ and $(i,j+1)$, 
\[ 
P_2 = \sum_{ij} \left[ \sum_{zz'} 
{\textstyle 
\left| 
	\begin{array} {cc} 
	1	&1 \\ 
	z	&z' 
	\end{array} 
\right\rangle 
\left \langle 
	\begin{array} {cc} 
	1	&1 \\ 
	z	&z' 
	\end{array} 
\right| 
+ 
\left| 
	\begin{array} {cc} 
	0	&0 \\ 
	1	&-1
	\end{array} 
\right\rangle 
\left \langle 
	\begin{array} {cc} 
	0	&0 \\ 
	1	&-1 
	\end{array} 
\right| 
} 
\right],
\] 
\[ 
P_3 = \sum_{ij} \left[ \sum_{z} 
{\textstyle 
\left| 
	\begin{array} {cc} 
	1	&0 \\ 
	z	&-1 
	\end{array} 
\right\rangle 
\left \langle 
	\begin{array} {cc} 
	1	&0 \\ 
	z	&-1
	\end{array} 
\right| 
+ 
\left| 
	\begin{array} {cc} 
	0	&1 \\ 
	1	&z 
	\end{array} 
\right\rangle 
\left \langle 
	\begin{array} {cc} 
	0	&1 \\ 
	1	&z 
	\end{array} 
\right| 
} 
\right]. 
\] 
The projector $P_4$ contains four-body interactions acting on the sites 
 $\{ (i,j-1), (i,j), (i,j+1), (i\pm1,j) \}$, 
\[ 
P_4 = 
\sum_{ij}  
 \sum_{z_1 z_2 z_3 z_4} 
 {\textstyle 
\left| 
	\begin{array}{cccc} 
	0	&0	&0	&1 \\ 
	z_1 	&z_2	&z_3	&z_4
	\end{array} 
\right \rangle 
\left \langle
	\begin{array}{cccc} 
	0	&0	&0	&1 \\ 
	z_1 	&z_2	&z_3	&z_4
	\end{array} 
\right |.
}  
\] 
The low energy subspace ($|\Psi_{\rm low} \rangle$) is defined by $P_{1,2,3,4} |\Psi_{\rm low} \rangle = 0$. 
The forbidden configurations by the above Hamiltonian projectors are illustrated in Fig.~\ref{fig:MBLErgodic} (b) ({\it main text}). 
The low-energy  quantum states in the subspace correspond to line configurations as shown in Fig.~\ref{fig:MBLErgodic} (a) ({\it main text}). 
We assume the strength of the energy penalty $\lambda$ is large enough to suppress all possible fluctuations leaving the subspace.

We now construct a microscopic local Hamiltonian that produces the effective Hamiltonian for low-energy quantum states in  Eq.~\eqref{eq:HamEff} ({\it main text}).  
The microscopic Hamiltonian takes a  form, 
\bea 
\label{eq:2dMicroHam} 
H = \sum_{i, j\in A} H_{ij}  ^A + \sum_{i,j\in B} H_{ij} ^B
\eea 
with $H_{ij} ^{A} $ and $H_{ij} ^B$  three-local interactions acting on the sites 
$\{(ij), (i+1,j),  (i,j+1) \} $, 
and $\{(i-1,j), (i,j), (i,j+1)\}$,  respectively, defined by  
\bea 
&&H_{ij}^{A} =
\sum_{z_1 z_2 }
 \textstyle  
 \left| 
 \begin{array}{ccc} 
 0      &1      &1 \\ 
 -1      &z_1    &z_2 
 \end{array} 
 \right\rangle 
  \left\langle 
 \begin{array}{ccc} 
 0      &1      &1 \\ 
 -1      &z_1    &z_2 
 \end{array} 
 \right|  \nn \\ 
&& {-\frac{1}{2}} 
  \sum_{z_{1,2}, z_{1,2}' } 
\textstyle \left[ U^{[ij]} _{z_1 z_2; z_1'z_2'}   
 \left| 
 \begin{array}{ccc} 
 0  &1  &1 \\ 
 -1  &z_2 &z_1 \\ 
 \end{array} \right\rangle 
 \left\langle 
 \begin{array}{ccc} 
 1      &1      &0 \\ 
 z_1'   &z_2'   &1 
 \end{array} 
 \right|
 +H.c. \right], \nn  \\ 
&&H_{ij}^{B}  = 
 \sum_{z_1 z_2 }
 \textstyle  
 \left| 
 \begin{array}{ccc} 
 1     &0      &1 \\ 
 z_1      &-1    &z_2 
 \end{array} 
 \right\rangle 
  \left\langle 
 \begin{array}{ccc} 
 1     	&0      &1 \\ 
 z_1     &-1    &z_2 
 \end{array} 
 \right| \nn \\
 && {-\frac{1}{2}} 
  \sum_{z_{1,2}, z_{1,2}' } 
\textstyle \left[  U^{[ij]} _{z_1 z_2; z_1'z_2'}   
 \left| 
 \begin{array}{ccc} 
 1  	&0  	&1 \\ 
 z_1  &-1 	&z_2 \\ 
 \end{array} \right\rangle 
 \left\langle 
 \begin{array}{ccc} 
 1      &1      &0 \\ 
 z_1'   &z_2'   &1 
 \end{array} 
 \right|
 +H.c. \right]. \nn  
\eea 
The interaction elements $U^{[ij]} _{z_1 z_2; z_1' z_2'} $ form a unitary matrix, with 
$\sum_{z_1 z_2} U^{[ij]} _{z_1' z_2' ; z_1 z_2} U^{[ij]*} _{z_1 z_2 ; z_1'' z_2''} =\delta_{z_1' z_1''} \delta_{z_2' z_2''}$. 
Projecting the microscopic Hamiltonian to the low-energy subspace, we reach the effective Hamiltonian provided in {\it the main text}, with the $\gamma$-states defined according to its wave function $\psi_l(\vec{z})$ by 
\[ 
\psi_{l_{ij}}  (\vec{z}) = \sum_{z_i' z_{i+1}'} U^{[ij]} _{z_i z_{i+1} ; z_i'z_{i+1}'} \psi_{l_{ij} -1} (z_0 \ldots z_{i-1}z_i' z_{i+1}' \ldots z_{I-1}) . 
\] 
The Hamiltonian is frustration free within the low-energy subspace. 
Taking  $l$ as a time step, the unitary update of $\psi_l(\vec{z})$ corresponds to unitary time evolution under a series of two-qubit gates defined by their matrix representation, $U^{[ij]}$.  
One explicit example can be found in Fig.~\ref{fig:suppfigHistoryStates}, the low energy subspace is spanned by the states illustrated by the line configurations. 
Our Hamiltonian construction has been inspired by the  geometrical Kitaev-Feynmann clock states, which have been used to prove the computational QMA-completeness of 2d quantum lattice models~\cite{2002_Kitaev_Book,2007_Lloyd_SIAM}.

\section{Mapping binary variable dynamics to Hamiltonian ground states}  
\label{sec:DynamicMapping} 
We consider a generic dynamical evolution of a binary sequence 
${\bf z}_t \equiv [z_{0,t}, z_{1, t}, \ldots, z_{I-1,t}]^T$ (a column vector), with $t$ the evolution time. 
A generic binary variable dynamics is described by a dynamical equation, 
\be 
\label{eq:generalevolve} 
{\bf z}_t = {\rm sgn} (F_t({\bf z}_{t-1})), 
\ee  
with $F_t$ an arbitrary function. 
A Turing complete  binary circuit can also be formulated in this form. 
The time-evolved configuration is the ground state of a static Hamiltonian, 
\be 
\label{eq:generalHam} 
H = -\sum_t {\bf z}_t ^T{\rm sgn} (F_t ({\bf z}_{t-1}))  , 
\ee 
by treating $t$ as a real space column index. The solution existence of Eq.~\eqref{eq:generalevolve} guarantees the Hamiltonian ground state is frustration free. 
If we further impose a local constraint on the binary variable dynamics, namely assuming $z_{i, t} $ is determined by its neighboring binary variables $z_{i'\in[i-M,i+M] , t-1}$ only ($M$ is a finite number), the  local dynamical equation takes a form 
\be 
z_{i,t} = {\rm sgn} \left[F_{i,t} (\{z_{i'\in [i-M, i+M], t-1}\} )\right].
\ee   
The Hamiltonian in Eq.~\eqref{eq:generalHam} is then reduced to 
\be 
H = -\sum_{i,t} z_{i,t} {\rm sgn} \left[ F_{i,t} (\{z_{i'\in [i-M, i+M], t-1}\} ) \right], 
\ee 
which only contains local couplings on a 2d grid. We thus conclude generic local Ising variable dynamics can be mapped to ground states of a local Hamiltonian. We expect this holds for arbitrary dynamical evolution in discrete settings, for that the discrete degrees of freedom can be rigorously  encoded by Ising variables.

\section{The exact ground state dual of time crystal}

In this section, we show that the non-equilibrium time crystal phase transition \cite{2012_Wilczek_TC,2012_Wilczek_QTC,2016_Sondhi_PRL,2016_Nayak_PRL,2017_Monroe_TC,2017_Lukin_Nature} also has an exact dual to a phase transition in the ground states of a static Hamiltonian. This Hamiltonian has the same form as that in Eq.~(6) in the main text,  
but with the $\gamma$-states 
$
|\gamma_l (\vec{z}_\star) \rangle = \sum_{\vec{z}} \psi_l( {\vec{z}} )  |l; \vec{z} \rangle, 
$
updated in a different way. Here for the time crystal, we update the wave functions $\psi_l(\vec{z})$ through a Floquet time crystal dynamics of a one-dimensional (1d) spin-$1/2$ system~\cite{2016_Nayak_PRL}. 
From $l=0$ to $l=I-1$, the update of $\psi_l$ corresponds to the unitary gates ${\rm e}^{{\rm i}\hat{x}_i h_x}$ ($i$ from $0$ to $I-1$), and then the unitary gates 
${\rm e}^{-{\rm i}\left(J^{[i]} \hat{z}_i\hat{z}_{i+1}+h_z ^{[i]} \hat{z}_i+h^{[i]}_x \hat{x}_i\right)}$ ($i$ from $I-1$ to $0$)
are applied from $l=I$ to $2I-1$. 
The amplitudes $J^{[i]}$, $h_z^{[i]}$, and $h_x ^{[i]}$ are chosen from certain random distributions. 
This unitary update process is then repeated forward for $J/2$ periods.

In Appendix~\ref{sec:QHam}, we have constructed a 2d local qudit Hamiltonian, whose projection on the low-energy subspace gives the effective Hamiltonian and whose ground state 
$
|G (\vec{z}_\star)  \rangle = \frac{1}{\sqrt{IJ}}  \sum_l | \gamma_l (\vec{z}_\star) \rangle
$ 
encodes the whole Floquet quantum dynamics. 
From the construction, we observe that this local qudit Hamiltonian has lattice translation symmetry (LTS) along the $j$-axis, 
with $A$ and $B$ sites forming a sublattice structure. 
Note that this LTS is dual to the time-translation symmetry (TTS) of the Floquet quantum dynamics. 
The time crystal phase transition arising from TTS breaking has an exact dual to the phase transition in the ground state $|G (\vec{z}_\star)  \rangle$ due to LTS breaking. The operator given in Eq.~(8) {\it in the main text} also characterizes this symmetry breaking. The corresponding order parameter reads 
\be
C(O) = \frac{1}{I} \left [ \sum_{i,j\in A} (-1)^{j /2} 
\langle G(\vec{z}_\star) | O_{ij} | G(\vec{z}_\star) \rangle  \right].
\ee
In the thermodynamic limit, this order parameter vanishes in the symmetric phase, and becomes finite in the LTS broken phase.

\bibliography{references}

\end{document}